\documentclass[12pt,a4paper]{article}

\usepackage{latexsym}
\usepackage{amssymb}
\usepackage{amsmath}
\usepackage[authoryear]{natbib}
\usepackage[latin1]{inputenc}

\newif\ifpdf
\ifx\pdfoutput\undefined
   \pdffalse
\else
   \pdfoutput=1
   \pdftrue
\fi

\ifpdf
    \usepackage{color}
    \definecolor{myred}{rgb}{0.5,0,0}
    \definecolor{myblue}{rgb}{0,0,0.75}
    \definecolor{mygreen}{rgb}{0,0.5,0}
   \usepackage[pdftex,
               pdftitle={A traffic lights approach to PD validation},
               pdfsubject={Validation},
               pdfkeywords={Binomial test, Vasicek model, granularity adjustment},
               pdfauthor={Dirk Tasche},
               pdfstartview=FitBH,
               breaklinks=true,
               colorlinks=true,
               citecolor=myred,
               linkcolor=myblue,
               urlcolor=mygreen]{hyperref}
\else
   \usepackage{nohyperref}
\fi

\begin{document}
\title{A traffic lights approach to PD validation%
}

\author{
Dirk Tasche\thanks{Deutsche Bundesbank, Postfach 10 06 02, 60006 Frankfurt am
Main, Germany\newline E-mail: tasche@ma.tum.de}\ %
\thanks{The opinions expressed in this note are those of the
author and do not necessarily reflect views shared by the Deutsche Bundesbank
or its staff.} }

\date{May 2, 2003}
\maketitle

\begin{abstract}
As a consequence of the dependence experienced in loan portfolios, the
standard binomial test which is based on the assumption of independence does
not appear appropriate for validating probabilities of default (PDs). The
model underlying the new rules for minimum capital requirements (Basle II) is
taken as a point of departure for deriving two parametric test procedures that
incorporate dependence effects.  The first one makes use of the so-called
granularity adjustment approach while the the second one is based on moment
matching.
\end{abstract}

The aim with this note is to present an approximate procedure for
one-observation-based inference on the adequacy of probability of default (PD)
forecasts. The PD forecast for a homogeneous portfolio of loans has to be
compared to the realized default rate one year later. In case of independent
default events, the natural procedure for this comparison would be the
standard binomial test. However, the well-known fact that default events are
correlated makes the binomial test appear unreliable.

Our approach here is to model the dependent default events in a Basle II-like
style and to arrive this way at a means to compute critical values which
respect correlations. However, in the Basle II-model the distribution of
default rates cannot be calculated with elementary arithmetic procedures as
they are available for instance in MSExcel. Therefore we suggest two
approximation schemes which seem to work with reasonable precision.

This note is organized as follows. In Section \ref{sec:1}, we describe the
general design of a traffic lights procedure for PD validation. Section
\ref{stoch} then specifies the stochastic model underlying the granularity
adjustment and moment matching approximation procedures to be introduced in
Sections \ref{sec:gran} and \ref{sec:mom} respectively. We conclude in Section
\ref{sec:num} with a numerical illustration of the approaches.

%%%%%%%%%%%%%%%
% New section %
%%%%%%%%%%%%%%%
\section{Setting the colors}
\label{sec:1}

We fix two \emph{confidence levels} $\alpha_{\mathrm{low}}$ and
$\alpha_{\mathrm{high}}$, e.g. $\alpha_{\mathrm{low}} = 95\%$ and
$\alpha_{\mathrm{high}} = 99.9\%$. Assume that the forecast for the default
rate is $p$. Under this assumption, we have to find \emph{critical values}
$c_{\mathrm{low}}$ and $c_{\mathrm{high}}$ such that the probabilities that
the realized number of defaults exceeds $c_{\mathrm{low}}$ and
$c_{\mathrm{high}}$ will equal $100\% - \alpha_{\mathrm{low}}$ and $100\% -
\alpha_{\mathrm{high}}$ respectively.

The traffic light for the adequacy of the PD forecast will be set
\emph{green}, if the realized number of defaults  is less than
$c_{\mathrm{low}}$. In this case there is no obvious contradiction between
forecast and realized default rate.

The traffic light will be set \emph{yellow}, if the realized number of
defaults is equal to or greater than $c_{\mathrm{low}}$ but less than
$c_{\mathrm{high}}$. The yellow light indicates that the realized default rate
is not compatible with the PD forecast. However, the difference of realized
rate and forecast is still in the range of usual statistical fluctuations. As
a consequence, the responsibility for the deviation of the forecast cannot
without doubt assigned to the portfolio manager.

The traffic light will be set \emph{red}, if the realized number of defaults
is equal to or greater than $c_{\mathrm{high}}$. In this case, the difference
of forecast and realized default rate is so large that any disbelief in a
wrong forecast would be unreasonable.

%%%%%%%%%%%%%%%
% New section %
%%%%%%%%%%%%%%%
\section{Specifying the stochastic model}
\label{stoch}

The first step towards determining the critical values $c_{\mathrm{low}}$ and
$c_{\mathrm{high}}$ is to fix a stochastic model that will enable us to carry
out the necessary numerical calculations. We take the Vasicek one factor-model
which underlies also the Basle II risk weight functions.

If $n$ denotes the number of loans in the portfolio under consideration and
$D_n$ is the realized number of defaults in the observed period of time, we
write $D_n$ as
\begin{equation}\label{eq:D_n}
  D_n\,=\,\sum_{i=1}^n \mathbf{1}_{\{\sqrt{\rho}\,X + \sqrt{1-\rho}\,\xi_i \le
  t\}}.
\end{equation}
In (\ref{eq:D_n}), $\mathbf{1}_E$ is the indicator function assuming the value
1 on the event $E$ and the value 0 on the complement of $E$. $X$ and $\xi_1,
\ldots, \xi_n$ are independent standard normal random variables. The threshold
$t$ has to be chosen in such a way that
\begin{subequations}
\begin{equation}\label{eq:E}
  \mathrm{E}[D_n]\,=\, n\,p.
\end{equation}
This will be achieved by setting
\begin{equation}\label{eq:t}
  t \, =\, \Phi^{-1}(p),
\end{equation}
\end{subequations}
with $\Phi$ denoting the standard normal distribution function. The choice of
the parameter $\rho$ (sometimes called \emph{asset correlation}) is not so
obvious. It should not be chosen higher than 0.24 which is the highest
correlation occurring in the Basle II rules. One way to handle this question
would be to leave the choice of a value for $\rho$ to the discretion of the
national supervisors. For instance, $\rho = 0.05$ appears to be appropriate
for Germany.

%%%%%%%%%%%%%%%
% New section %
%%%%%%%%%%%%%%%
\section{The granularity adjustment approach}
\label{sec:gran}

In order to make work the traffic lights approach we have to find methods for
calculating the critical values which have been introduced in Section
\ref{sec:1}. For instance, the critical value $c_{\mathrm{low}}$ is
characterized by
\begin{subequations}
\begin{equation}\label{eq:c_low}
c_{\mathrm{low}}\,=\,\min\{ k:\,\mathrm{P}[D_n \ge k]\le
1-\alpha_{\mathrm{low}} \}.
\end{equation}
(\ref{eq:c_low}) is equivalent to
\begin{equation}\label{eq:cquant}
c_{\mathrm{low}}\,=\,q(\alpha_{\mathrm{low}}, D_n)+1,
\end{equation}
where $q(\alpha, D_n)$ denotes the usual $\alpha$-quantile of $D_n$, i.e.
\begin{equation}\label{eq:quant}
q(\alpha, D_n) \,=\, \min\{x: \mathrm{P}[D_n \le x]\ge \alpha\}.
\end{equation}
Analogously, we have
\begin{equation}\label{eq:cquant_high}
  c_{\mathrm{high}}\,=\,q(\alpha_{\mathrm{high}}, D_n)+1.
\end{equation}
\end{subequations}
Write
\begin{subequations}
\begin{equation}\label{eq:R_n}
  R_n \,=\, D_n/n
\end{equation}
for the \emph{default rate} corresponding to the number of defaults $D_n$. For
computational reasons, it is often appropriate to consider $R_n$ instead of
$D_n$. But, of course, the quantiles of $R_n$ and $D_n$ are related by
\begin{equation}\label{eq:RDquants}
  n\,q(\alpha, R_n) \,=\, q(\alpha, D_n).
\end{equation}
\end{subequations}
Since the distribution of $D_n$ cannot be calculated with elementary methods,
\citet{Gordy01} suggested the \emph{granularity adjustment} approach for
approximating the quantiles $q(\alpha, R_n)$. This approximation is based on a
second order Taylor expansion of $q(\alpha, R_n)$ in the following sense
\begin{subequations}
\begin{align}\label{eq:taylor}
  q(\alpha, R_n)
   & =\, q(\alpha, R + h\,(R_n-R))\Big|_{h=1}\\
  & \approx\,
  q(\alpha, R) + \frac{\partial}{\partial h} \,q(\alpha, R + h\,(R_n-R))\Big|_{h=0} +
  \frac1 2\,\frac{\partial^2}{\partial h^2} \,q(\alpha, R+ h\,(R_n-R))\Big|_{h=0}.\notag
\end{align}
The random variable $R$ in (\ref{eq:taylor}) can be chosen as
\begin{equation}\label{eq:lim}
  R\,=\,\lim_{n\to\infty} R_n
  \,=\,\Phi\bigl(\frac{t-\sqrt{\rho}\,X}{\sqrt{1-\rho}}\bigr).
\end{equation}
The quantile $q(\alpha, R)$ turns out to be
\begin{equation}\label{eq:qD}
  q(\alpha,R)\,=\,\Phi\bigl(\frac{\sqrt{\rho}\,\Phi^{-1}(\alpha)+t}{\sqrt{1-\rho}}\bigr),
\end{equation}
and, as a consequence, can  easily be calculated.
\end{subequations}
Unfortunately, when defining $R$ with (\ref{eq:lim}), the partial derivatives
in (\ref{eq:taylor}) may not exist because the distribution of $D_n$ is purely
discrete. However, \citet{MartinWilde02} noted that although (\ref{eq:taylor})
was derived for \emph{smooth} distributions its application may yield sensible
results even in semi-smooth situations. Using the formulas for the derivatives
by \citet{MartinWilde02} one arrives at \citep[cf.][]{Tasche03}
\begin{align}\label{eq:gran}
 q(\alpha,D_n)  & \approx n\,q(\alpha, R) + \frac 1{2} \bigg(2\,q(\alpha,R) - 1  \\
   & \qquad + \frac{q(\alpha,R)\,(1-q(\alpha,R))}{\phi\bigl(
   \frac{\sqrt{\rho}\,q(1-\alpha,X)-t}{\sqrt{1-\rho}}\bigr)}
   \left(\frac{\sqrt{\rho}\,q(1-\alpha,X)-t}{\sqrt{1-\rho}} -
   \sqrt{\frac{1-\rho}{\rho}}\,q(1-\alpha,X)\right)\bigg),\notag
\end{align}
where $\phi(x) = (2\,\pi)^{-1}\,e^{- x^2/2}$ denotes the standard normal
density.

For given forecasted PD $p$ and asset correlation $\rho$, by means of
(\ref{eq:cquant}), (\ref{eq:cquant_high}), and (\ref{eq:gran}) the critical
values for the traffic lights approach respecting correlation can be
calculated.

%%%%%%%%%%%%%%%
% New section %
%%%%%%%%%%%%%%%
\section{Fitting the default rate distribution with moment matching}
\label{sec:mom}

A further approach to determine the critical values defined by
(\ref{eq:cquant}) and (\ref{eq:cquant_high}) can be based on approximating the
distribution of $R_n$ as given by (\ref{eq:D_n}) and (\ref{eq:R_n}) with a
Beta-distribution. When proceeding this way, the parameters of the
Beta-distribution are determined by matching the expectation and the variance
of $R_n$ \citep[cf.][]{Over00}.

Recall that the density of a $B(a,b)$-distributed random variable $Z$ is
defined by
\begin{subequations}
\begin{align}\label{eq:dens}
  \beta(a,b;x) & = \frac{\Gamma(a+b)}{\Gamma(a)\,\Gamma(b)}\,x^{a-1}\,(1-x)^{b-1}, \ 0 < x < 1, \\
  \intertext{where $\Gamma$ denotes the Gamma-function expanding the factorial
  function to the positive reals, and that the expectation and the variance
  respectively of $Z$  are given by}
  \mathrm{E}[Z] & = \frac a{a+b},\label{eq:exp}\\
  \mathrm{var}[Z] & = \frac{a\,b}{(a+b)^2\,(a+b+1)}.\label{eq_var}
\end{align}
\end{subequations}
Equating the right-hand sides of (\ref{eq:exp}) and (\ref{eq_var}) with
$\mathrm{E}[R_n]$ and $\mathrm{var}[R_n]$ respectively leads to
\begin{subequations}
\begin{align}\label{eq:a}
  a & = \frac{\mathrm{E}[R_n]}{\mathrm{var}[R_n]}\,\bigl(\mathrm{E}[R_n]\,
  (1-\mathrm{E}[R_n])-\mathrm{var}[R_n]\bigr) \\
  \intertext{and}
  b & = \frac{1-\mathrm{E}[R_n]}{\mathrm{var}[R_n]}\,\bigl(\mathrm{E}[R_n]\,
  (1-\mathrm{E}[R_n])-\mathrm{var}[R_n]\bigr).\label{eq:b}
\end{align}
\end{subequations}
It is not hard to show that
\begin{subequations}
\begin{align}\label{eq:p}
  \mathrm{E}[R_n] & = p \\
  \intertext{and}
  \mathrm{var}[R_n] & = \frac {n-1}n\,\Phi_2(t,t,\rho) + \frac p n -
  p^2,\label{eq:Phi2}
\end{align}
with $t$ defined by (\ref{eq:t}) and $\Phi_2$ denoting the bivariate standard
normal distribution function.
\end{subequations}
Since common tools like MSExcel have not got implemented algorithms for the
calculation of $\Phi_2$, the approximation
\begin{equation}\label{eq:Phi}
\Phi_2(t,t,\rho)\,\approx\, \Phi(t)^2 + \frac{e^{-t^2}}{2\,\pi}\,(\rho +
1/2\,\rho^2\,t^2)
\end{equation}
can be used in (\ref{eq:Phi2}). (\ref{eq:Phi}) is based on a second order
Taylor expansion of $\Phi_2(t,t,\rho)$ with respect to $\rho$
\citep[cf.][]{Tong90} and yields a fairly good approximation for moderate
values of $\rho$ and $\alpha$. As will be shown in Section \ref{sec:num}, the
quality of approximation decreases for $\rho \ge 0.2$ and values of $\alpha$
close to one.

By means of (\ref{eq:Phi}), (\ref{eq:p}), and (\ref{eq:Phi2}), the quantile
$q(\alpha, D_n)$ can be calculated approximately via
\begin{equation}\label{eq:approx}
  q(\alpha, D_n) \,\approx\,n\,q(\alpha, Z),
\end{equation}
where $Z$ is a Beta-distributed random variable and the parameters $a$ and $b$
of its distributions are given by (\ref{eq:a}) and (\ref{eq:b}) respectively.

%%%%%%%%%%%%%%%
% New section %
%%%%%%%%%%%%%%%
\section{Numerical examples}
\label{sec:num}

For the purpose of illustration of the previous sections, we
calculated\footnote{Upon request, an MSExcel-sheet with implementations of the
algorithms can be obtained from the author.} the lower and higher critical
values of the traffic lights for various portfolio sizes $n$ and two different
asset correlation values. Table~1 shows the 95\%-critical values for a test of
$\text{PD} \le 0.01$ in case of low ($\rho = 0.05$) and high ($\rho = 0.2$)
asset correlation. We compare the results obtained with the classical binomial
test, the granularity adjustment of Section \ref{sec:gran}, and the moment
matching of Section \ref{sec:mom}.
\begin{table}[h]
\begin{center}
\begin{tabular}{|l|c|c|c||c|c|c|}  \hline
  % after \\ : \hline or \cline{col1-col2} \cline{col3-col4} ...
  \multicolumn{1}{|c|}{$n$} & 50 & 250 & 1000 & 50 & 250 & 1000\\ \hline
  Approach: & \multicolumn{3}{c||}{$\rho = 0.05$}& \multicolumn{3}{c|}{$\rho = 0.2$}\\ \hline
  Binomial & 2 &  5 &  15 & 2 &  5 &  15\\ \hline
  Granularity adjustment & 3 &  7 &  24 & 3 &  11 &  39\\ \hline
Moment matching & 4 &  8 &  25 & 4 &  12 &  42\\ \hline
\end{tabular}
{\small\sl \caption{95\%-critical values for PD tests of hypothesis $p \le
0.01$.} }
\end{center}
\label{tb:1}
\end{table}
Since the binomial test relies on the assumption of independent default
events, there is no difference when switching the correlation regime from low
to high. For the other approaches, this picture changes dramatically.
Respecting the correlation makes grow a lot the critical values, with more
than doubling in case of high correlation. At this quite moderate confidence
level of 95\%, the results of the granularity adjustment and the moment
matching approach do not differ much, although differences appear to become
larger with an increasing portfolio size and higher correlation.

As the numbers in Table~2 show, the effects described for the case of a
moderate confidence level become much more pronounced when we consider a very
high confidence level like 99.9\%.
\begin{table}[h]
\begin{center}
\begin{tabular}{|l|c|c|c||c|c|c|} \hline
  % after \\ : \hline or \cline{col1-col2} \cline{col3-col4} ...
  \multicolumn{1}{|c|}{$n$} & 50 & 250 & 1000 & 50 & 250 & 1000\\ \hline
  Approach: & \multicolumn{3}{c||}{$\rho = 0.05$}& \multicolumn{3}{c|}{$\rho = 0.2$}\\ \hline
  Binomial & 4 &  9 &  21 & 4 &  9 &  21 \\ \hline
  Granularity adjustment & 6 &  15 &  50 & 9 &  38 &  148\\ \hline
Moment matching & 7 &  16 &  47 & 10 &  33 &  118\\ \hline
\end{tabular}
{\small\sl \caption{99.9\%-critical values for PD tests of hypothesis $p \le
0.01$.} }
\end{center}
\label{tb:2}
\end{table}
In particular, in case of high correlation the differences between the results
with the granularity approach and the moment matching approach respectively
cannot be any longer neglected. Since the granularity adjustment is based on
an approximation procedure in the tail of the loss distribution, whereas the
moment matching comes from an approximation in the center of the distribution,
the granularity adjustment results will in general be the more reliable. But
this observation should not be seen as a knock-out criterion against the
moment matching since the use of high correlations like 0.2 might be
considered being too conservative.

%%%%%%%%%%%%%%%
% New section %
%%%%%%%%%%%%%%%

\end{document}